\documentclass[11pt,a4paper,english]{IEEEtran}
\usepackage{mathptmx}

\usepackage[T1]{fontenc}
\usepackage[latin9]{inputenc}
\usepackage{units}
\usepackage{amstext}
\usepackage{graphicx}
\PassOptionsToPackage{normalem}{ulem}
\usepackage{ulem}

\makeatletter

\pdfpageheight\paperheight
\pdfpagewidth\paperwidth

\usepackage{cite}

\makeatother

\usepackage{babel}
\begin{document}

\title{Towards a Circular Economy via Intelligent Metamaterials}

\author{Christos Liaskos, Ageliki Tsioliaridou, Sotiris Ioannidis\\
{\small{}Foundation for Research and Technology - Hellas (FORTH)}\\
{\small{}Emails: \{cliaskos,atsiolia,sotiris\}@ics.forth.gr}}
\maketitle
\begin{abstract}
The present study proposes the use of intelligent metasurfaces in
the design of products, as enforcers of circular economy principles.
Intelligent metasurfaces can tune their physical properties (electromagnetic,
acoustic, mechanical) by receiving software commands. When incorporated
within products and spaces they can mitigate the resource waste caused
by inefficient, partially optimized designs and security concerns.
Thus, circular economy and fast-paced product design become compatible.
The study begins by considering electromagnetic metamaterials, and
proposes a complete methodology for their deployment. Finally, it
is shown that the same principles can be extended to the control of
mechanical properties of objects, exemplary enabling the micro-management
of vibrations and heat, with unprecedented circular economy potential.
\end{abstract}


\section{Introduction\label{sec:Introduction}}

The industrial and electronic revolution had a tremendous impact on
our world at all levels, from micro to macro. Our every-day lives
are depended on the facilities of industrial production of goods,
transportation, communication and computation. At macroscopic level,
the global economy is shaped by the flow of goods and services. However,
this revolution was partial and never planned in depth, to account
for its ecological sustainability. As a result, the natural resources
of the planet are already expended faster than their replenishment
rate, while by 2030 the expenditure will be twice the replenishment~\cite{oecdobserver}.

Industrial design has been partial due to its fast-paced, antagonistic
nature and due to the lack of a central framework for life-cycle management
across products. The fast-paced nature means that a company that offers
a product to the market is income-driven, and always insecure about
another company getting to the market first. Thus, due to lack of
time, products are only partially optimized: they revolve about the
immediate facility, disregarding all the rest\textendash sustainability
and security included. This fast pace also alter the environment faster
than the reaction time of governments. Legislation and frameworks
arrive late, while their enforcement is slow as well.

The concept of circular economy seeks to provide a framework for sustainability
at micro and macro levels~\cite{murray2017circular}. At micro level
(single product), it provides directives for a circular product life-cycle.
Instead of the traditional, linear order of life-cycles phases, i.e.,
i) raw resource acquirement, ii) processing, iii) distribution, iv)
use and v) disposal, circular economy creates links from disposal
to all preceding phases. According to it, product design should facilitate
\uline{four} links: i) decomposition to raw materials, ii) re-processing
or refurbishment, iii) re-deployment and redistribution, and iv) multiple
uses. At macro-level (in a horizontal, cross-product approach), it
is expected to provide a legal framework for enforcing these approaches.
However, this effort will require extra design and development time
from companies, which opposes their fast-paced nature. Thus, novel
approaches are required to ensure the success of the circular economy
concept.

The present work proposes the use of intelligent metamaterials as
enforcers of circular economy in a fast-paced product design. Metamaterials
are artificial materials with engineered physical properties (such
as electromagnetic-EM, acoustic and mechanical behavior). Moreover,
they can exhibit properties not found in natural materials, such as
negative refraction index, perfect insulation, etc. Additionally,
a recent advancement called HyperSurfaces offers metamaterials with
software-defined properties~\cite{LiaskosCACM2018}. A well-defined
programming interface abstracts underlying complexities and allows
non-specialists (such as software developers) to incorporate the HyperSurface
in applications and products, without caring for their inner physics.

Due to these traits, HyperSurfaces and intelligent metasurfaces in
general can act as mitigation agents for partially optimized product
designs. We envision metasurfaces as coatings or structural parts
of products. For instance, electromagnetic interference and unwanted
emissions can be harvested by HyperSurface-coated walls and be transformed
back to usable electrical or mechanical energy~\cite{LiaskosWOWOMOM2018,LLiaskosWCM2018,LiaskosCACM2018}.
Mechanical metamaterials can micro-manage emanated heat and vibrations
from motors to recycle it as energy while effectively cooling it.
Acoustic metamaterials can surround noisy devices or applies on windows
to provide a more silent environment, but to also harvest energy which
can be added to the household. A notable trait of metamaterials is
that they are simple structures and, therefore, their production can
easily follow the aforementioned four principles of circular economy.
Moreover, their software-defined nature allows for quickly ``patching''
the non-ecological parts of new or existing products, without much
overhead to the industrial pace. The ``patching'' may also be deferred
in the form of ``eco-firmware'', distributed via the Internet to
ecologically tune a single product or horizontal sets of products.

In the following, we provide a methodology creating ecosystems of
intelligent metamaterials and propose the micro-Managed Electromagnetic
Environments. We begin by focusing on EM metamaterials (Sections \ref{sec:Enabling-Technologies}-\ref{subsec:Section-B.-Methodology}),
since they are more well-known to the general audience and, thus,
easier to describe. Lastly, we generalize to the case of mechanical
and acoustic metamaterials (Section~\ref{sec:Extending-to-Micro-Management}).

\section{Enabling Technologies\label{sec:Enabling-Technologies}}

\subsection{EM Metamaterials and Metasurfaces \label{sec:Metasurfaces}}

This section provides the necessary background knowledge on metamaterials
metasurfaces, discussing dimensions and composition, operating principles,
programmable metasurfaces and supported functionalities.

A regular metasurface is a planar, artificial structure which comprises
a repeated element, the meta-atom, over a substrate. In most usual
compositions, the meta-atom is conductive and the substrate is dielectric.
Common choices are copper over silicon, printed organic circuits over
films, while silver and gold constitute conductors for exotic applications~\cite{Zhu.2017}.
Other approaches employ graphene, in order to interact with $THz$-modulated
waves~\cite{Lee.2012}. Metasurfaces are able to control EM waves
impinging on them, in a frequency span that depends on the overall
dimensions. The size of the meta-atom is comparable to the intended
interaction wavelength,~$\lambda$, with $\nicefrac{\lambda}{2}$
constituting a common choice. The thickness of the metasurface is
smaller than the interaction wavelength, ranging between $\nicefrac{\lambda}{10}\to\nicefrac{\lambda}{5}$
as a rule of a thumb. Metasurfaces usually comprise several hundreds
of meta-atoms, which results into fine-grained control over the EM
interaction control. Metamaterials are the 3D counterpart of the concept,
and can be perceived as a stack of metasurfaces.
\begin{figure}[t]
\begin{centering}
\includegraphics[width=1\columnwidth]{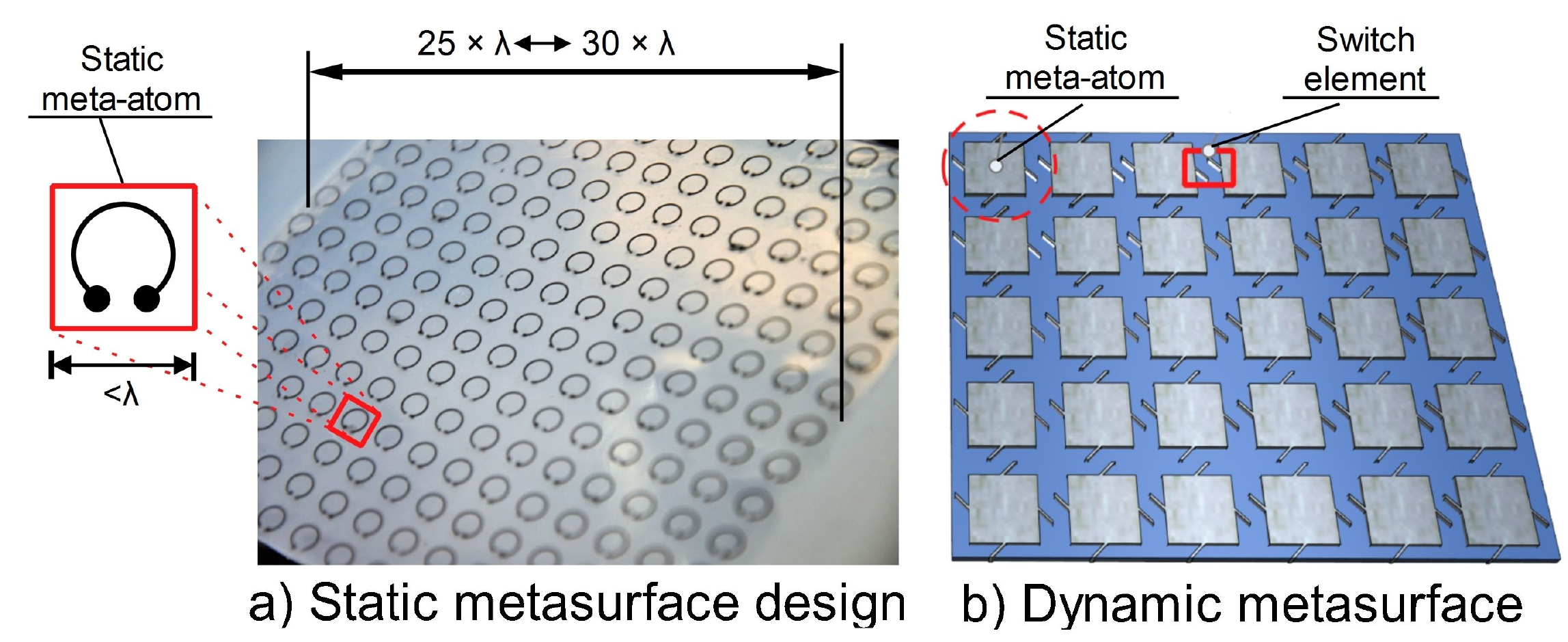}
\par\end{centering}
\caption{\label{fig:Mspatterns}Split ring resonators (left) constituted a
very common type of static metasurfaces, with fixed EM behavior. Dynamic
designs (right) incorporate switch elements (MEMS, CMOS or other)
to offer dynamically tunable EM behavior.}
\end{figure}

Figure~\ref{fig:Mspatterns}-a illustrates a well-studied metasurface
design comprising split-ring resonators as the meta-atom pattern.
Such classic designs that rely on a static meta-atom, naturally yield
a static interaction with EM waves. The need for dynamic alteration
of the EM wave control type has given rise to dynamic, programmable
metasurfaces, exemplary illustrated in Fig.~\ref{fig:Mspatterns}-b.
Dynamic meta-atoms incorporate phase switching components, such as
MEMS or CMOS transistors, which can alter the structure of the meta-atom.
Thus, dynamic meta-atoms allow for time-variant EM interaction, while
meta-atom alterations may give rise to multi-frequency operation~\cite{Zhu.2017}.
Phase switching components can also be classified into state-preserving
or not. For instance, mechanical switches may retain their state and
require powering only for state transitions, while semiconductor switches
require power to maintain their state.

\begin{figure}[t]
\begin{centering}
\includegraphics[width=1\columnwidth]{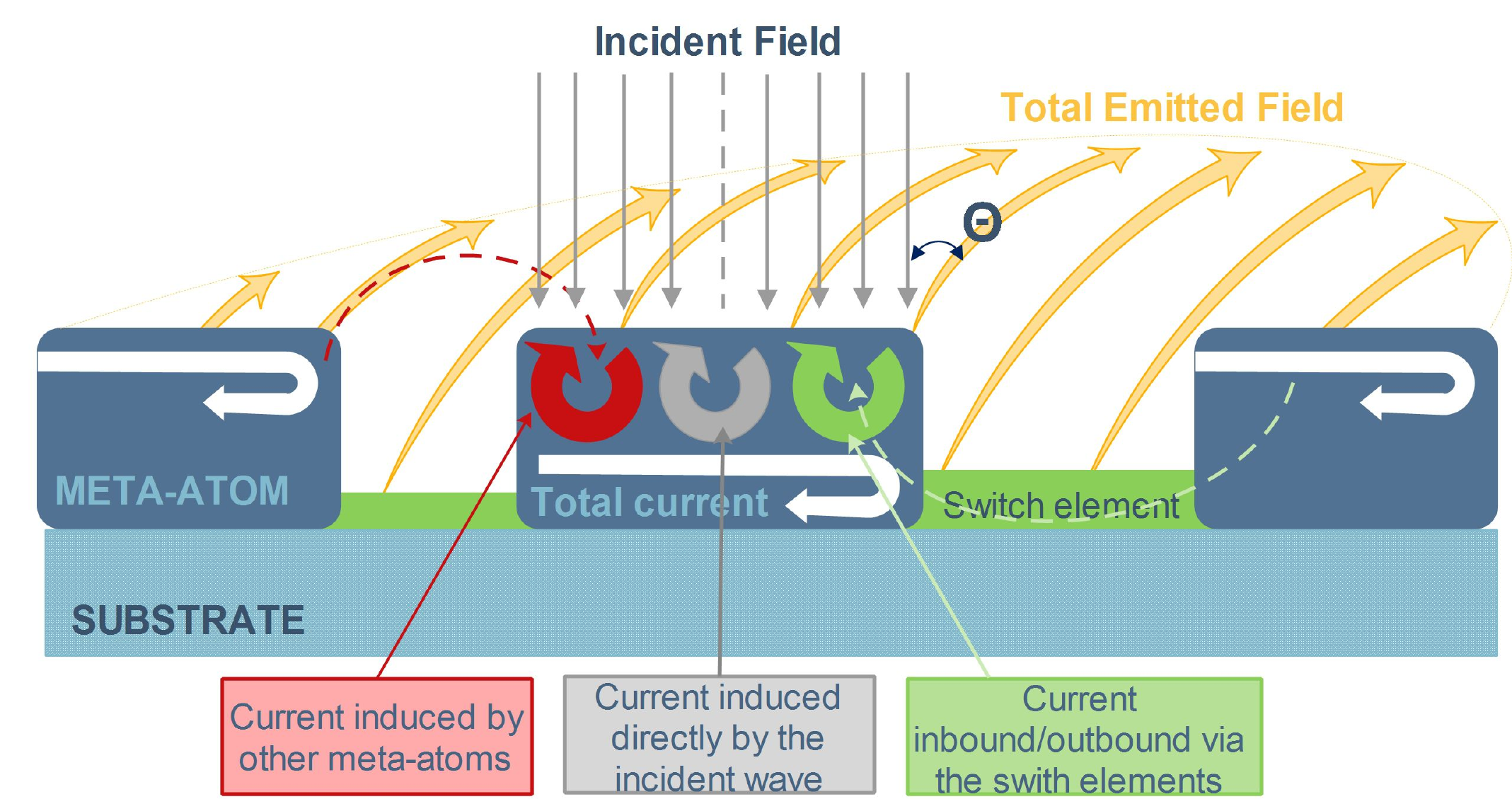}
\par\end{centering}
\caption{\label{fig:MSprinciple}The principle of metasurface functionality.
Incident waves create a well-defined EM response to the unit cells.
The cell response is crafted in such a way that the aggregate field
follows a metasurface-wide design objective, e.g., reflection towards
a custom angle $\Theta$.}
\end{figure}

The operating principle of metasurfaces is given in Fig.~\ref{fig:MSprinciple}.
The meta-atoms, and their interconnected switch elements in the dynamic
case, act as control factors over the surface currents flowing over
the metasurface. The total EM response of the metasurface is then
derived as the total emitted field by all surface currents, and can
take completely engineered forms, such as the unnatural reflection
angle shown in Fig.~\ref{fig:MSprinciple}. Engineering the total
surface current is a complex process that must account for currents
directly induced over the metasurface by the incident wave, the currents
induced in a meta-atom wirelessly by other meta-atoms, as well as
the currents flowing inwards or outwards from a meta-atom via the
switch elements.

A metasurface can support a wide range of EM interactions, i.e., \emph{EM}
\emph{functions}, which are classified as follows~\cite{Minovich.2015}:
\begin{itemize}
\item Reflection of an impinging wave, with a given direction of arrival,
towards a completely custom direction.
\item Refraction of EM waves via the metasurface towards any inwards direction.
Both the reflection and refraction functions can override the outgoing
directions predicted by Snell's law. Reflection and refraction functions
will jointly be referred to as wave \emph{steering}. Steering can
provide security at physical layer, by bending waves around unwanted
users, completely negating eavesdropping~\cite{LLiaskosWCM2018}.
\item Wave absorbing, i.e., ensuring minimal reflected and/or refracted
power for impinging waves.
\item Wave polarizing, i.e., changing the oscillation orientation of the
wave's electric and magnetic field.
\end{itemize}
In general, metasurfaces and metamaterials can fully re-engineer the
impinging wave, producing a completely custom response-field~\cite{Lucyszyn.2010}.
HyperSurfaces are an incorporation-ready approach to metasurfaces~\cite{LiaskosWOWOMOM2018,LiaskosCACM2018,LLiaskosWCM2018},
which provide both the hardware and software to control their EM properties.
They model the EM functions as a software API, comprising callbacks
with the following general form:
\[
{\scriptstyle \texttt{outcome}\gets\texttt{callback(action\_type, parameters)}}
\]
where $\texttt{action\_type}$ defines the type of EM function (such
as $\texttt{STEER}$, $\texttt{ABSORB}$, etc.), accompanied by the
necessary parameters to completely define the wanted interaction.
Thus, the complexities of metasurface Physics are abstracted and hidden,
focusing on usability by siftware developers.

\subsection{The Internet of (Nano)-Things\label{sec:IoT}}

The Internet of Things (IoT) is a rapidly growing ecosystem of sensory
and communication platforms. IoT devices commonly employ  inexpensive
sensors and microprocessors, tiny power supplies and wireless transceivers,
prioritizing scalability, robust communication and energy efficiency.
Via IoT, the number of connected devices increases rapidly and extends
to the control over any ordinary object in our environment, such as
switches, medical implants, cars, trackers, clothes, lights and door
locks. Novel IoT products are being released almost daily, at a trend
that is expected to yield 20-30 billion connected IoT devices by 2020~\cite{Nordrum.2016}.
IoT platforms are promising choices for ambient environmental monitoring
and control.

Moreover, ongoing IoT research now targets the nanoscale, with the
target to achieve ultimate levels of minification, scalability and
energy-efficiency. Research has started shrinking sensors, antennas
transceivers and logic circuits to the nanoscale~\cite{Akyildiz.2010c},
small enough to circulate within living bodies and to mix into industrial
materials, taking medicine, energy efficiency and many construction
sectors to a new level. Protocols able to sustain the vast device
numbers and sensory data volumes are under active research, with highly
promising results~\cite{Tsioliaridou.2017,Tsioliaridou.2015,Tsioliaridou.2016}.
Due to their minimal size and vast numbers, nanosensors can gather
information from a multitude of environmental views. External data
aggregators, e.g., in the Cloud, can then generate incredibly detailed
snapshots containing the slightest changes in light, vibration, electrical
currents and magnetic fields, for uniquely-detailed environmental
monitoring.

\subsection{Network Function Virtualization and Software-Defined Networking \label{sec:SDN_NFV}}

Network functions virtualization (NFV) is a novel concept in network
design and operation, which seeks to express services offered by a
network into building-block components that can be connected (i.e.,
\emph{chained}) together, to create custom operations on demand.

In essence, NFV is a form of well-structured virtualization that abstracts
an offered function from the underlying \emph{material} performing
the low-level computations. In its classic form, NFV building blocks\textendash the
Virtual Network Function Components\textendash are distributed as
well-isolated packages that can be configured, initiated, migrated
and destroyed in accordance with a defined workflow or life-cycle.
The NFV principles do not specify a strict format for the Components,
but rather guidelines on how to effectively structure them to be chain-able
in a scalable manner. The Components themselves and the virtualization
approach followed to contain and distribute them can be completely
custom and heterogeneous. Moreover, the applications of NFV span across
domains, and can be employed in a variety of settings~\cite{Nandugudi.2016}.

Software-defined networking (SDN) is a novel way of managing networks
that has gained significant traction in the industrial and the academic
world~\cite{Henneke.2016,NunesBrunoAstutoA.2014}. Its core principles
are: i) to well-separate the network control from the network data,
ii) to provide a clean interface for interacting with the control,
and iii) to provide a central view of the various distributed forwarding
elements. This is accomplished by delegating network control decisions
to a central service, which has a bird's eye view of the controlled
system, and configures its operation in response to policies and events.

In this study, the SDN and NFV are used as abstract approaches for
organizing software in general, rather than in their literal use in
the narrow field of computer networking.

\section{A Methodology for Micro-Managed Electromagnetic Environments\label{subsec:Section-B.-Methodology}}

\begin{figure*}[t]
\begin{centering}
\includegraphics[width=0.9\textwidth]{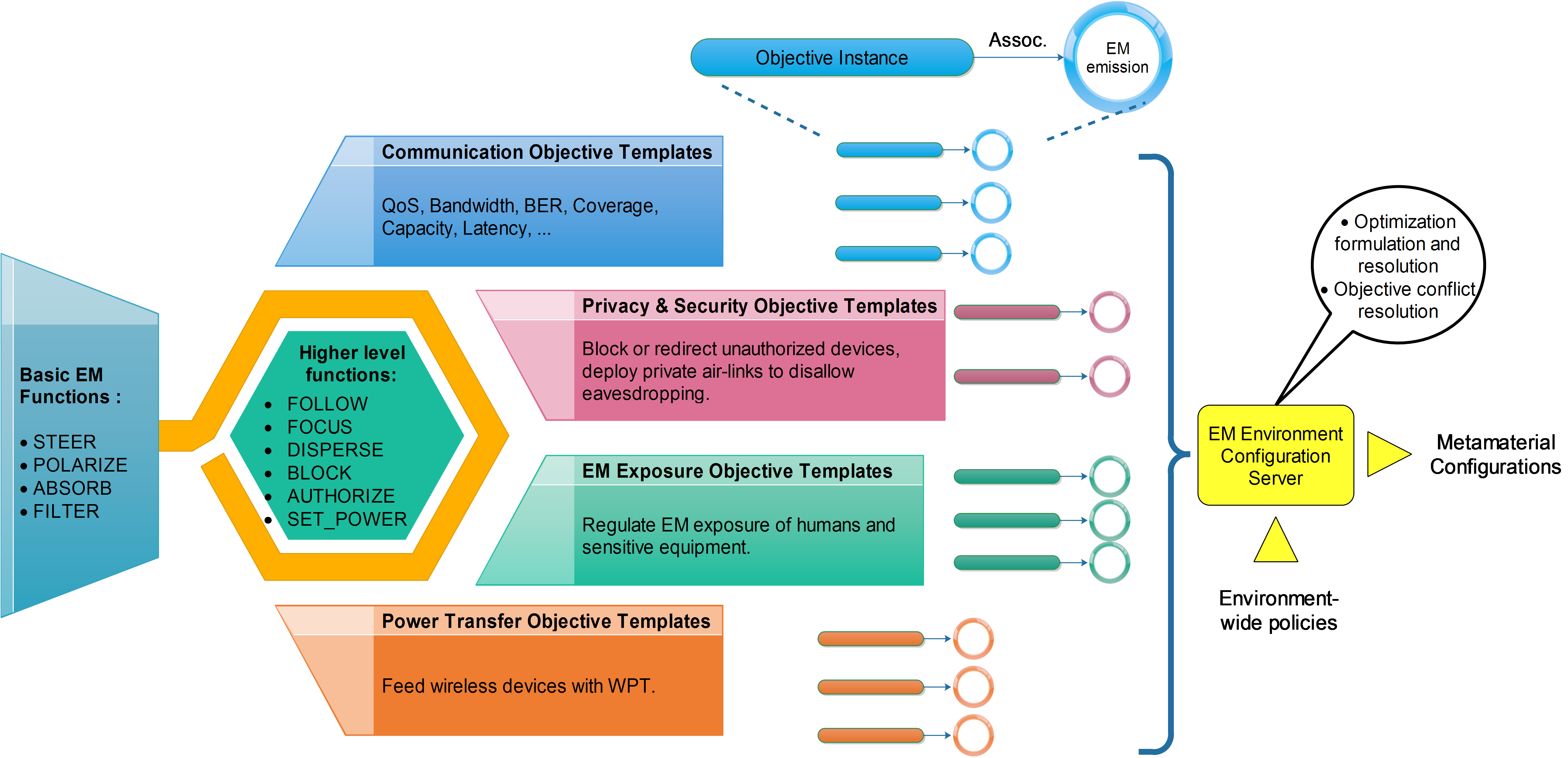}
\par\end{centering}
\caption{\label{fig:Workflow}Workflow overview of the proposed Micro-Managed
Electromagnetic Environments (mMEEs).}
\end{figure*}
\begin{figure}[t]
\begin{centering}
\includegraphics[width=1\columnwidth]{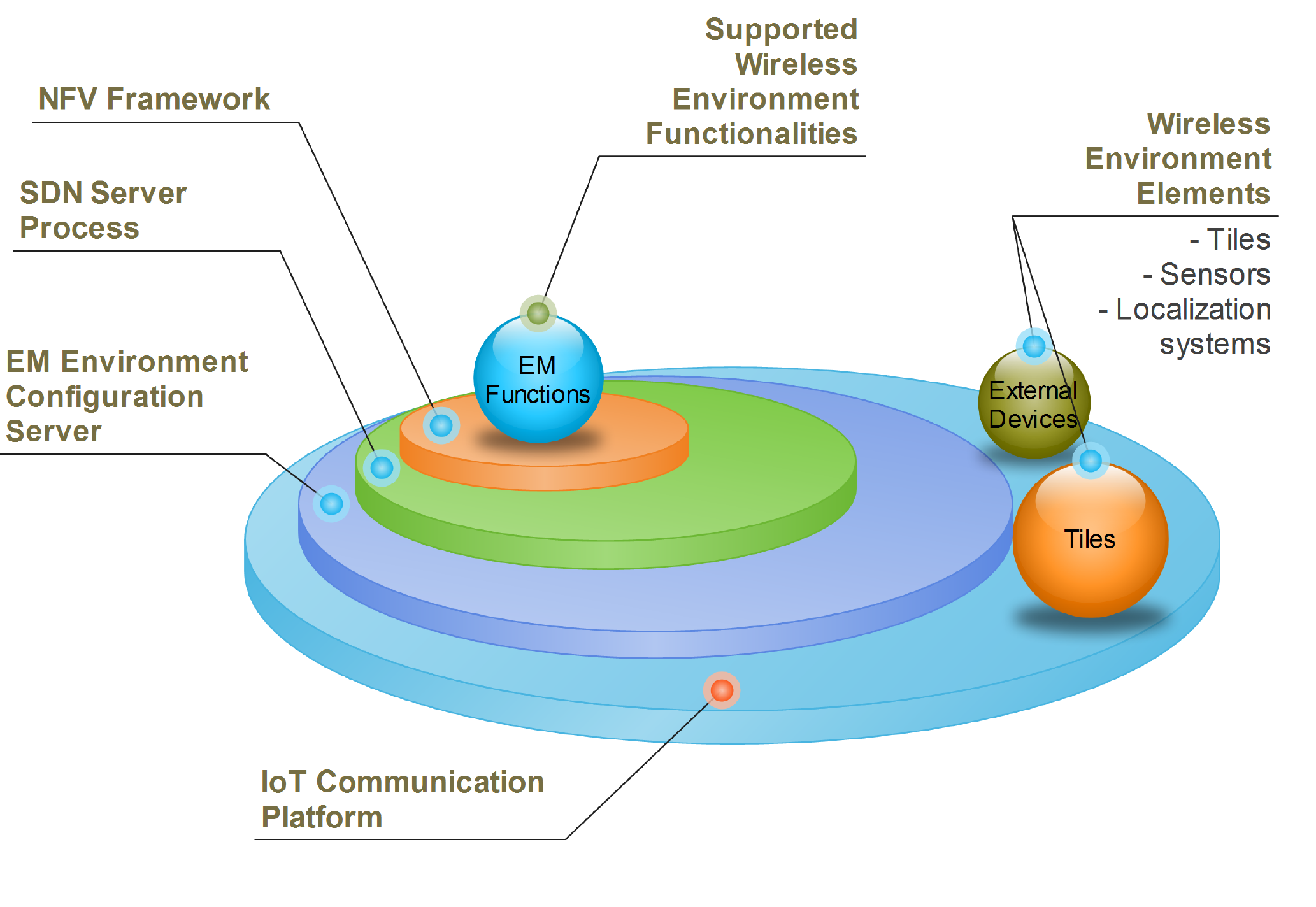}
\par\end{centering}
\caption{\label{fig:WhyMakeSense}Incorporation schematic of the involved enabling
technologies, to form Micro-Managed Electromagnetic Environments.}
\end{figure}

Figures \ref{fig:Workflow} and \ref{fig:WhyMakeSense} provide an
overview of design methodology for Micro-Managed Electromagnetic Environments
(mMEEs). Figure~\ref{fig:Workflow} illustrates the envisioned workflow
of mMEEs, while Fig.~\ref{fig:WhyMakeSense} depicts the incorporation
scheme of the enabling technologies.

The mMEE design methodology incorporates three core ideas:
\begin{enumerate}
\item Express the EM functions of tiles within a NFV framework. Exploit
the modularity and chaining capabilities of the NFV paradigm, and
allow for a well-organized hierarchy of EM functionalities that can
be used in a wide application context.
\item Implement the expressed EM virtual functions in an SDN-over-IoT platform.
Make use of the centralized control offered by SDN, to obtain the
requirements of the wireless devices present within a space. SDN and
IoT combined facilitate the interfacing with other systems (e.g.,
localization) and the control over the numerous tiles within the environment.
\item Employ optimization principles to combine and host multiple EM virtual
function instances within an mMEE. Express the wireless device requirements
in the form of well-defined optimization objectives, facilitating
their ``compilation'' to matching tile configurations.
\end{enumerate}
A driving principle within the methodology is to make mMEE accessible
to a broad developer audience, without requiring expertise in Physics,
following the paradigm set by HyperSurfaces. Therefore, the envisioned
hierarchy of EM functionalities follows the levels shown in Fig.~\ref{fig:Workflow}.

At the lowest level, we envision wave-front manipulation EM functions,
which will allow for custom steering, polarization, absorption and
non-linear (e.g., frequency-selective) filtering of waves impinging
upon tiles. This basic set of functions will be chained together towards
the formation of intermediate-level functionalities. As an example,
we propose functions such as $\texttt{FOCUS}$ and $\texttt{FOLLOW}$.
The former refers to a lens functionality, aimed at gathering ambient
energy and directing it towards a wireless device for remote charging.
The latter makes an EM function applicable to moving targets. Thus,
$\texttt{FOCUS}$ and $\texttt{FOLLOW}$ could be combined to mitigate
path loss for a mobile device within the environment. Additional examples
of higher-level functions include the $\texttt{DISPERSE}$, for achieving
wireless coverage within a space, $\texttt{BLOCK}$ (opposite to $\texttt{AUTHORIZE}$),
for absorbing transmissions from malevolent or unauthorized users
before they propagate within the environment, and $\texttt{SET\_POWER}$
to define the total power level carried towards a device or a space,
e.g., for wireless power transfer (WPT) tasks.

At the final level, the hierarchy exposes functionality objective
templates, which are envisioned as well-defined and parametric optimization
goals, organized into four, broad application contexts:
\begin{itemize}
\item Communication objectives, which aim to provide advanced QoS to wireless
devices, leveraging the mMEE potential. These exemplary include bandwidth
maximization, Bit Error Rate minimization, extended high-quality wireless
coverage and cross-device interference negation.
\item Privacy and Security objectives, which aim at blocking eavesdropping
altogether via ``private air-paths'' and the isolation of unauthorized
or malevolent transmissions within a space.
\item EM Exposure objectives, which aim at keeping the exposure of humans
and sensitive equipment (e.g., medical) within acceptable levels.
\item Wireless Power Transfer objectives, aiming at supplying power to properly
equipped wireless devices (e.g., supporting energy harvesting).
\end{itemize}
An SDN control service is the focal point for the mMEE operation.
In collaboration with localization and environmental sensing systems,
it gathers information on the present wireless devices, their configurations
and connectivity intentions. Subsequently, it expresses these intentions
to corresponding objective instances, i.e., a parameterized association
between an objective template and a set of wireless emissions.

We envision environment-wide policies, apart from device-specific
objectives. These policies will allow for the expression of:
\begin{enumerate}
\item General limits that should be respected within the wireless environment.
For instance, a maximum allowed value of wireless power throughout
the environment. We note that this is an additional feature enabled
by mMEEs, which cannot be supported by plain environments.
\item mMEE resource reservations, e.g., the ``air-paths'' that should
be kept spare as slack reserve, ensuring the proper and timely handling
of emergency requests for EM micro-management.
\item Sub-space optimizations, such as offering a specific EM micro-management
type within a sub-space only.
\item Priorities for the resolution of conflicting user requirements. For
instance, a user may require WPT within an area of strict EM silence.
\end{enumerate}
The SDN controller passes the knowledge of: i) the needs of the wireless
devices present, and ii) the environment policies, to an optimization
service. This service deduces the mMEE configuration that best fit
the objectives, subject to the policy restrictions. The SDN controller
and the optimization service will be engaged in an online control
loop, constantly matching the currently gathered knowledge to the
optimal mMEE response.

We note that the nature of the mMEE objectives and policies is mathematical,
as they correspond to EM profiles (e.g., desired ranges of EM field
values) that should be present at points within a space (device locations).
On the other hand, the ``inputs'' to the optimization problem are
the supported EM functions that can be deployed to each tile, which
can be viewed as ``discrete variables''. Therefore, optimization
techniques based on Mixed-Integer Programming~\cite{Dantzig.2016}
can constitute a promising resolution direction. Metaheuristic optimization
approaches, e.g., Swarm-based or Genetic algorithms, offer an alternative
direction, with multiple success stories across disciplines~\cite{Luke.2009}.
Finally, we highlight the research direction of artificial intelligence-controlled
mMEEs, where Neural Networks can be trained for a given space, i.e.,
learning the necessary tile EM profiles that correspond to time-variant
user needs~\cite{Su.2017}. This direction could exemplary study
the mMEE configuration for filtering WiFi signals by high-level attributes
such as their SSID, deliberately scrambling signals and making them
readable only within a given room, and more.

\section{Extending to Micro-Management of Mechanical Properties\label{sec:Extending-to-Micro-Management}}

EM metamaterials were the first kind of metamaterials to be proposed
and studied, mainly due to the relative ease of manufacturing. However,
the same principles have been applied to control sound (acoustic metamaterials~\cite{acoustic_meta})
and mechanical waves (mechancical metamaterials~\cite{mechanic_meta}).
With the advent of 3D printing, acoustic and mechanical metamaterials
have become easier to manufacture boosting the related research.

Acoustic metamaterials can manipulate and re-engineer sound waves
in gases, solids and liquids. This control is exerted mainly through
the bulk modulus, mass density and chirality. The latter two properties
correspond to the electromagnetic permittivity and permeability in
EM metamaterials. Related to this are the mechanics of wave propagation
in a lattice structure. Also materials have mass, and instrinsic degrees
of stiffness. Together these form a resonant system, and the sonic
resonance is excited by sonic frequencies (e.g., pulses at audio frequencies).
Controlling sonic waves has been extended to unnatural properties,
including negative refraction.

Mechanical metamaterials can be seen as a superset of acoustic metamaterials.
They too can be designed to exhibit properties which cannot be found
in nature. Popular mechanical properties that have been controlled
in academic studies include compressibility, contractivity and focusing
of mechanical waves. However, the exerted control over vibrations
can customized as required by the application scenario.

The functionalities of acoustic and mechanical metamaterials can be
exposed in software, following the HyperSurface example. Subsequently,
and due to the identical principles of operation, they can be incorporated
to a micro-management environment. The later can be derived by generalizing
the presented mMEE.

\section{Conclusion\label{sec:Conclusion}}

The present paper introduced the use of intelligent metasurfaces and
metamaterials, as enforcers of circular economy principles in the
life-cycle of products. The study exploits the fact that intelligent
metamaterials can change the electromagnetic, mechanical or acoustic
properties following software directives. These artificial materials
can be incorporated within products, objects and spaces and micro-manage
electrical and mechanical energy. Thus, they can mitigate on-the-fly
the ecologic discrepancies of the original product design. The study
presented a methodology for organizing, controlling and orchestrating
intelligent materials, while discussing their potential in circular
economics.

\section*{Acknowledgment}

This work was partially funded by the European Union via the Horizon
2020: Future Emerging Topics call (FETOPEN), grant EU736876, project
VISORSURF (http://www.visorsurf.eu).

\bibliographystyle{IEEEtran}

\end{document}